\documentclass[3p,times,procedia]{elsarticle}
\usepackage{nupha_ecrc}
\usepackage{graphicx}

\volume{00}

\firstpage{1}

\journalname{Nuclear Physics A}

\runauth{}

\jid{nupha}

\jnltitlelogo{Nuclear Physics A}

\usepackage{amssymb}

\usepackage[figuresright]{rotating}

\newcommand {\jpsi}{J/\psi}
\newcommand{\X}{\chi_{c1}(3872)}

\newcommand\BR {\mathcal{B}}

\begin{document}

\begin{frontmatter}



\dochead{XXVIIIth International Conference on Ultrarelativistic Nucleus-Nucleus Collisions\\ (Quark Matter 2019)}

\title{LHCb measurements of the exotic tetraquark candidate $\chi_{c1}(3872)$ in high-multiplicity $pp$ and $p$Pb collisions}


\author{J. Matthew Durham for the LHCb collaboration}

\address{Los Alamos National Laboratory, Los Alamos, NM 87545 USA}

\begin{abstract}

The last decade of hadron spectroscopy has unveiled a wealth of states that do not have the properties expected of particles composed of two or three valence quarks. Among the most intriguing of these exotics is the $\chi_{c1}(3872)$, which various models attempt to describe as a hadronic molecule, a compact tetraquark, an unexpected charmonium state, or their mixtures. To date, most experimental studies of the $\chi_{c1}(3872)$ have focused on its production through $B$ meson decays. Heavy ion collisions, as well as high multiplicity $pp$ collisions, offer a new window on the properties of this poorly understood hadron. In these systems, promptly produced $\chi_{c1}(3872)$ hadrons can interact with other particles in the nucleus and/or those produced in the collision. The influence of these interactions on the observed $\chi_{c1}(3872)$ yields provides information that can help discriminate between the various models of its structure, as well as give insight into the dynamics of the bulk particles produced in these collisions.  With a full range of precision vertexing, tracking, and particle ID capabilities covering 2 to 5 in units of rapidity, the LHCb experiment is especially well suited to measurements of both prompt and non-prompt exotic hadrons. These proceedings present new LHCb measurements $\chi_{c1}(3872)$ production in high multiplicity $pp$ collisions and $p$Pb collisions through the decay to $J/\psi \pi^{+} \pi^{-}$.

\end{abstract}


\end{frontmatter}


\section{Introduction}

Bound states of more than three valence quarks have been expected since the proposal of the quark model of hadron structure more than 50 years ago \cite{GellMann, Zweig}.  However, unambiguous evidence for these exotic multiquark states has only recently become available via data from $B$ factories and the Large Hadron Collider (LHC).  More than twenty exotic states containing pairs of heavy charm or bottom quarks have been identified \cite{ExoticReview}, none of which are predicted to exist by potential models which successfully describe the observed conventional quarkonia states \cite{CharmoniumPotentialModel}.  

The most well-studied exotic hadron is the $\X$.  First discovered by the Belle experiment in 2003 \cite{Belle_X3872obs}, it has since been confirmed by multiple other experiments \cite{CDF_X3872obs, BABAR_X3872obs, LHCb_X3872obs}.  However, despite intense experimental and theoretical study, the exact nature of the $\X$ is not clear. Some models have described it as a compact tetraquark \cite{X3872asDiquark, X3872asDiquark2, X3872asTetraquark}, which is relatively tightly bound and has a radius of ${\sim}$1 fm.  The proximity of the $\X$ mass and the sum of the $D^{0}$ and $\overline{D}$$^{*0}$ masses led to models of the $\X$ as a $D^{0}$$\overline{D}$$^{*0}$ hadronic molecule \cite{X3872asDD,X3872asDD2}, with a small binding energy and, consequently, a large radius of ${\sim}7$ fm.  Mixtures of various exotic and conventional states have also been considered \cite{X3872asMix}.

The dense QCD environment produced in heavy ion collisions and high-multiplicity $pp$ collisions provides a new method for examining the properties of the $\X$.  In these collisions, $\X$ hadrons that are produced promptly at the collision vertex can undergo interactions with other particles produced in the event, which can lead to breakup.  The magnitude of this effect is expected to be dependent on the binding energy of the $\X$, as has been suspected for breakup of $\jpsi$ and $\psi$(2S) in heavy ion collisions \cite{Ferreiro_comovers}.  Therefore, measurements of prompt $\X$ production in high multiplicity collisions may provide data that can discriminate between tightly-bound tetraquark and weakly-bound hadronic molecule scenarios.  

\section{Measurement}

\begin{figure}[h]
  \centering
 \includegraphics[width=0.5\linewidth]{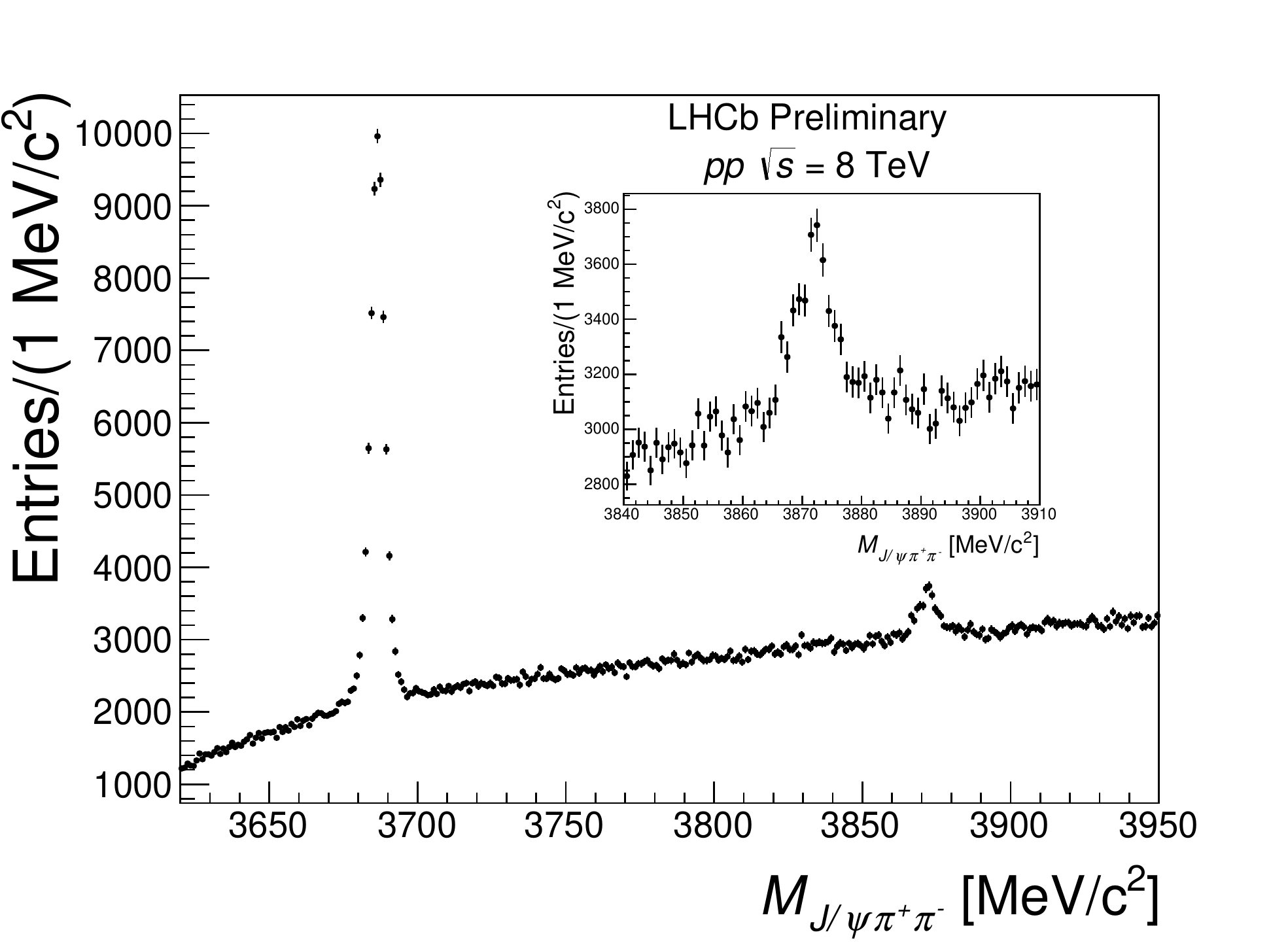}
  \caption{The $\jpsi \pi^+ \pi^-$ invariant-mass spectrum.  The inset shows the region of the $\X$ resonance.}
  \label{fig:mass}
\end{figure}

The LHCb detector is a single-arm forward spectrometer covering the pseudorapidity range $2 < \eta < 5$, described in detail in Ref.~\cite{LHCb-DP-2008-001}.  In this analysis, $\jpsi$ candidates that are produced in $pp$ collisions at $\sqrt{s}$ = 8 TeV are identified through their decay to $\mu^{+}\mu^{-}$.  These candidates are combined with opposite-sign pairs of charged pions, and a kinematic refit of the tracks is performed which constrains the $\jpsi$ mass to the known value and requires all four tracks to originate from precisely the same vertex.  The resulting $\jpsi\pi^{+}\pi^{-}$ invariant mass spectrum shown in Fig. \ref{fig:mass}, where peaks corresponding to the $\psi$(2S) and $\X$ are clearly visible.  From this invariant mass spectrum, a direct comparison between the conventional quarkonium state $\psi$(2S) and the exotic $\X$ can be made.

At the LHC, both the $\psi$(2S) and the $\X$ can be produced promptly at the collision vertex, or in the decays of hadrons containing $b$ quarks.  Taking advantage of the relatively long lifetime of $B$ hadrons, the prompt and $b$-decay contributions are separated via a simultaneous fit to the invariant mass and proper time spectrum of $\psi$(2S) and $\X$ candidates (see Ref. \cite{LHCb-CONF-2019-005} for details).  The fit returns the fraction of inclusive hadrons that are produced promptly at the collision vertex, $f_{prompt}$.

To examine the multiplicity dependence of $\psi$(2S) and $\X$ production, the data is separated into bins of event activity, which is quantified by the number of tracks reconstructed in the LHCb VELO detector $N_{tracks}^{VELO}$.  This serves as a metric for the number of charged particles produced in the event.

\section{Results}

\begin{figure}
\centering
\begin{minipage}{0.5\textwidth}
  \centering
  \includegraphics[width=1\linewidth]{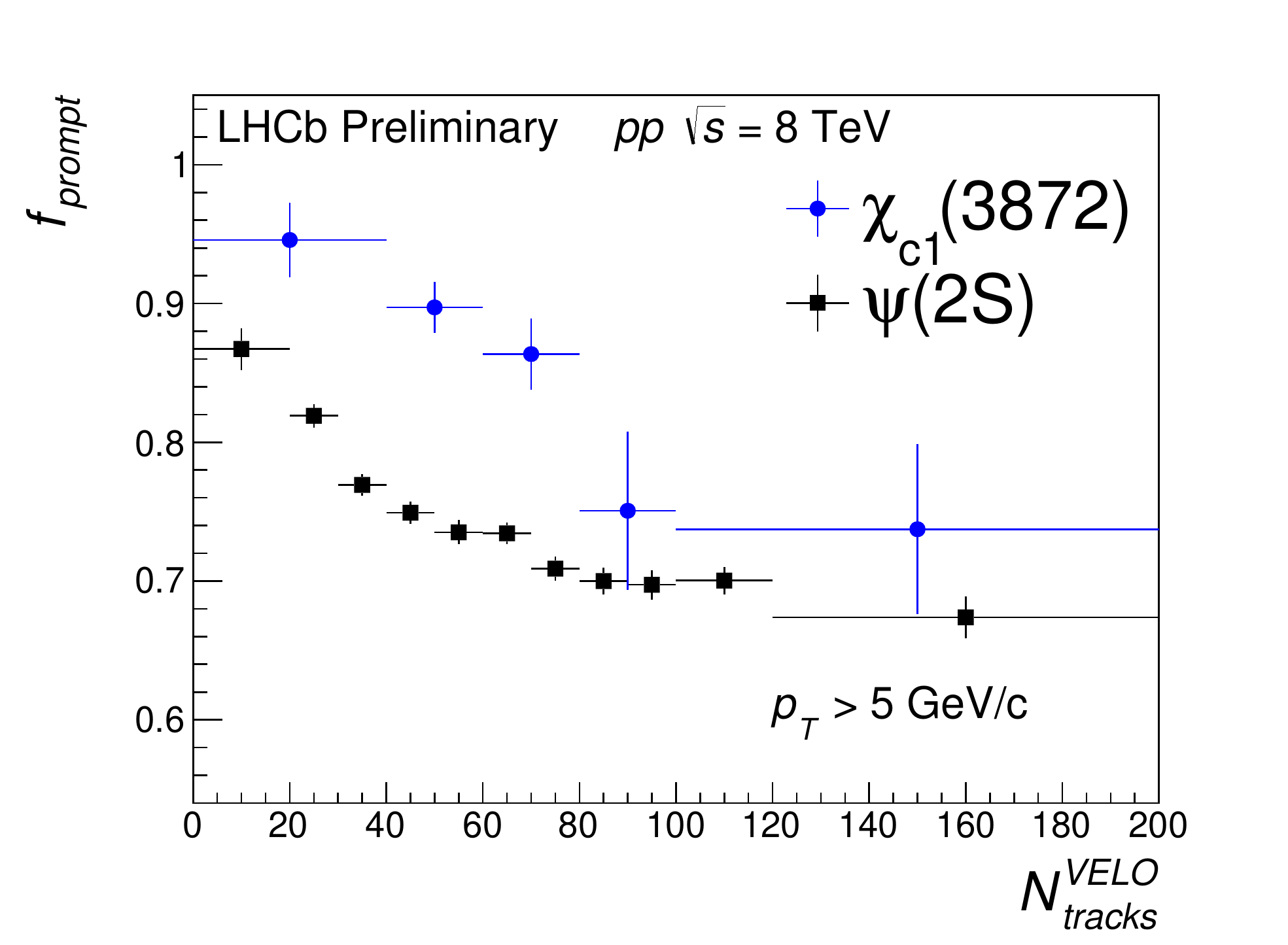}
  \label{fig:test1}
\end{minipage}%
\begin{minipage}{0.5\textwidth}
  \centering
  \includegraphics[width=1\linewidth]{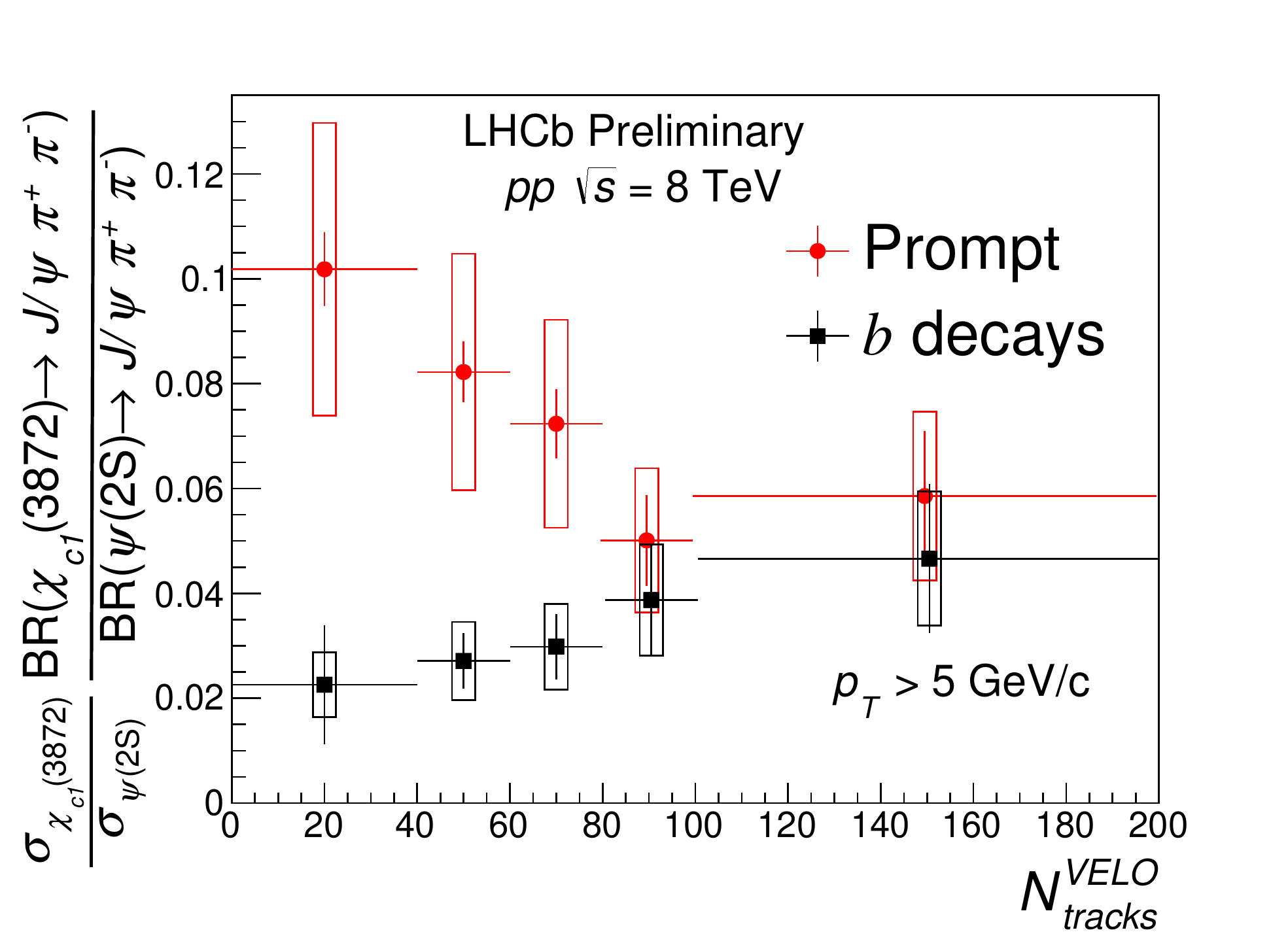}
  \label{fig:test2}
\end{minipage}
\caption{Left: the fraction of $\psi$(2S) and $\X$ hadrons produced promptly at the $pp$ collision vertex.  Right: the ratio of $\X$ and $\psi$(2S) cross sections measured in the $\jpsi \pi^{+} \pi^{-}$ channel as a function of the number of tracks reconstructed in the VELO. }
\label{fig:Fig2}
\end{figure}

The prompt fraction $f_{prompt}$ of $\psi$(2S) and $\X$ hadrons is shown in the left panel of Fig. \ref{fig:Fig2}.  We see that the ratio decreases with increasing charged particle multiplicity for both species, which may be due to a combination of effects.  Events in which a pair of $b$ quarks are produced, hadronize, and decay will naturally have a higher multiplicity than events without $b$ quarks, which will drive $f_{prompt}$ down as multiplicity increases.  An additional effect which may come into play is the suppression of promptly produced $\psi$(2S) and $\X$ hadrons, which can be disrupted through interactions with other particles produced in the event.  This will also decrease $f_{prompt}$ with increasing multiplicity.

To examine these prompt and $b$ decay components separately, we calculate the ratio of cross sections $\sigma_{\X}/\sigma_{\psi(2S)}$ times the respective branching fractions to $\jpsi \pi^+ \pi^-$,

\begin{equation}
\frac{\sigma_{\X}}{\sigma_{\psi(2S)}} \frac{\BR[\X \rightarrow \jpsi \pi^+\pi^-]}{\BR[\psi(2S) \rightarrow \jpsi \pi^+\pi^-]} = \frac{N_{\X} \,f_{prompt}^{\X}}{N_{\psi(2S)} \, f_{prompt}^{\psi(2S)}}\frac{\varepsilon_{\psi(2S)}}{\varepsilon_{\X}}
.
\end{equation}

Here $N$ is the inclusive signal yield, $f_{prompt}$ is the prompt fraction, and $\epsilon$ represents the efficiency for reconstructing and selecting the hadron of interest.  This ratio is shown for both prompt and $b$ decay production in the right panel of Fig. \ref{fig:Fig2}.  Here the error bars corresponds to uncertainties that are point-to-point uncorrelated, which are dominated by the statistical uncertainties on the prompt fraction of the $\X$ sample, and the boxes are fully correlated systematic uncertainties that are dominated by the efficiency corrections.  We see that the ratio of prompt cross sections decreases with increasing multiplicity, as is expected in a scenario where weakly-bound $\X$ hadrons are being disrupted more than the conventional $\psi$(2S) state via interactions with comoving particles.  The ratio of cross sections from $b$ decays, which are produced away from the primary collision vertex in vaccuum, does not show any significant dependence on multiplicity.  This is expected, as this ratio is set by the branching fractions of $B$ hadron decays to $\psi$(2S) and $\X$, and is not affected by event activity at the primary vertex.

\section{Summary and Outlook}
We have found that the fraction of both $\psi$(2S) and $\X$ which are produced promptly at the collision vertex decreases with increasing charged particle multiplicity in $pp$ collisions at $\sqrt{s}$ = 8 TeV.  The ratio of the prompt cross sections $\sigma_{\X}/\sigma_{\psi(2S)}$ also decreases with multiplicity, while the ratio of cross sections from decays of $B$ hadrons remains constant within uncertainties.  This could indicate that promptly produced $\psi$(2S) and $\X$ hadrons are being broken up via interactions with other particles produced in the event.  These suppression more significantly affects the exotic $\X$ than the conventional $\psi$(2S), which may indicate that the $\X$ has a smaller binding energy than the $\psi$(2S).  In this case, the $\X$ may be a very weakly bound state, such as a hadronic molecule.

\begin{figure}[h]
  \centering
\includegraphics[width=0.51\linewidth, angle = 270,trim={0 0 4cm 0},clip]{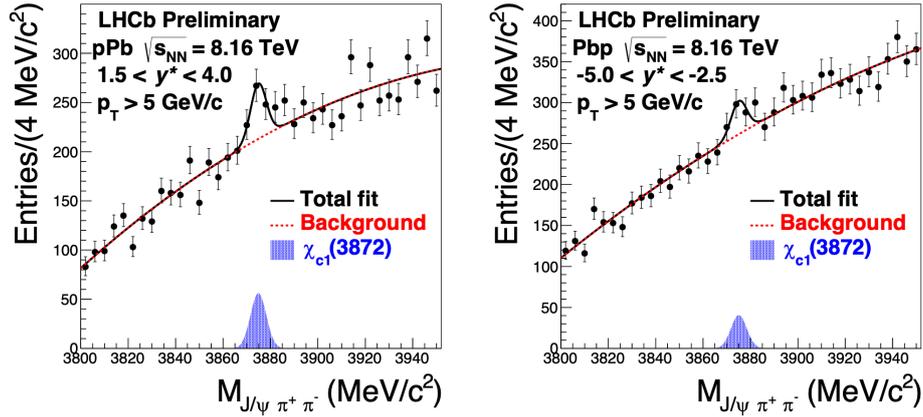}
  \caption{The $\jpsi \pi^+ \pi^-$ invariant-mass spectrum in $p$Pb (left) and Pb$p$ (right) collisions at $\sqrt{s_{NN}}$ = 8.16 TeV.}
  \label{fig:pPb}
\end{figure}

LHCb is also pursuing measurements of $\X$ production in $p$Pb collisions, see Fig. \ref{fig:pPb}.  Forthcoming results on the nuclear modification factor $R_{pPb}$ for $\X$ will provide additional constraints on the structure of this poorly understood exotic hadron.

\bibliographystyle{elsarticle-num}
\bibliography{nupha-template.bib}

\begin{thebibliography}{10}
\expandafter\ifx\csname url\endcsname\relax
  \def\url#1{\texttt{#1}}\fi
\expandafter\ifx\csname urlprefix\endcsname\relax\def\urlprefix{URL }\fi
\expandafter\ifx\csname href\endcsname\relax
  \def\href#1#2{#2} \def\path#1{#1}\fi

\bibitem{GellMann}
M.~Gell-Mann, A schematic model of baryons and mesons, Physics Letters 8~(3)
  (1964) 214 -- 215.
\newblock \href
  {http://dx.doi.org/https://doi.org/10.1016/S0031-9163(64)92001-3}
  {\path{doi:https://doi.org/10.1016/S0031-9163(64)92001-3}}.

\bibitem{Zweig}
{G. Zweig, CERN Report No. CERN-TH-401}, {An SU(3) model for strong interaction
  symmetry and its breaking}, 1964.

\bibitem{ExoticReview}
S.~L. Olsen, T.~Skwarnicki, D.~Zieminska, {Nonstandard heavy mesons and
  baryons: experimental evidence}, Rev. Mod. Phys. 90 (2018) 015003.
\newblock \href {http://dx.doi.org/10.1103/RevModPhys.90.015003}
  {\path{doi:10.1103/RevModPhys.90.015003}}.

\bibitem{CharmoniumPotentialModel}
T.~Barnes, S.~Godfrey, E.~S. Swanson, Higher charmonia, Phys. Rev. D 72 (2005)
  054026.
\newblock \href {http://dx.doi.org/10.1103/PhysRevD.72.054026}
  {\path{doi:10.1103/PhysRevD.72.054026}}.

\bibitem{Belle_X3872obs}
S.-K. Choi, et~al., {Observation of a narrow charmoniumlike state in exclusive
  $B^\pm\ensuremath{\rightarrow}K^\pm
  {\ensuremath{\pi}}^{+}{\ensuremath{\pi}}^{\ensuremath{-}}J/\ensuremath{\psi}$
  decays}, Phys. Rev. Lett. 91 (2003) 262001.
\newblock \href {http://dx.doi.org/10.1103/PhysRevLett.91.262001}
  {\path{doi:10.1103/PhysRevLett.91.262001}}.

\bibitem{CDF_X3872obs}
D.~Acosta, et~al., {Observation of the narrow state
  $X(3872)\ensuremath{\rightarrow}J/\ensuremath{\psi}{\ensuremath{\pi}}^{+}{\ensuremath{\pi}}^{\ensuremath{-}}$
  in $p\overline{p}$ collisions at $\sqrt{s}=1.96$ TeV}, Phys. Rev. Lett. 93
  (2004) 072001.
\newblock \href {http://dx.doi.org/10.1103/PhysRevLett.93.072001}
  {\path{doi:10.1103/PhysRevLett.93.072001}}.

\bibitem{BABAR_X3872obs}
B.~Aubert, et~al., {Study of the
  ${B}^{\ensuremath{-}}\ensuremath{\rightarrow}J/\ensuremath{\psi}{K}^{\ensuremath{-}}{\ensuremath{\pi}}^{+}{\ensuremath{\pi}}^{\ensuremath{-}}$
  decay and measurement of the
  ${B}^{\ensuremath{-}}\ensuremath{\rightarrow}X(3872){K}^{\ensuremath{-}}$
  branching fraction}, Phys. Rev. D 71 (2005) 071103.
\newblock \href {http://dx.doi.org/10.1103/PhysRevD.71.071103}
  {\path{doi:10.1103/PhysRevD.71.071103}}.

\bibitem{LHCb_X3872obs}
R.~Aaij, et~al., {Observation of $X(3872) $ production in $pp$ collisions at
  $\sqrt{s} = 7$ TeV}, Eur. Phys. J. C72 (2012) 1972.
\newblock \href {http://arxiv.org/abs/1112.5310} {\path{arXiv:1112.5310}},
  \href {http://dx.doi.org/10.1140/epjc/s10052-012-1972-7}
  {\path{doi:10.1140/epjc/s10052-012-1972-7}}.

\bibitem{X3872asDiquark}
L.~Maiani, F.~Piccinini, A.~D. Polosa, V.~Riquer, {Diquark-antidiquark states
  with hidden or open charm and the nature of $X(3872)$}, Phys. Rev. D 71
  (2005) 014028.
\newblock \href {http://dx.doi.org/10.1103/PhysRevD.71.014028}
  {\path{doi:10.1103/PhysRevD.71.014028}}.

\bibitem{X3872asDiquark2}
R.~D. Matheus, S.~Narison, M.~Nielsen, J.-M. Richard, {Can the $X(3872)$ be a
  ${1}^{++}$ four-quark state?}, Phys. Rev. D 75 (2007) 014005.
\newblock \href {http://dx.doi.org/10.1103/PhysRevD.75.014005}
  {\path{doi:10.1103/PhysRevD.75.014005}}.

\bibitem{X3872asTetraquark}
S.~Dubnicka, A.~Z. Dubnickova, M.~A. Ivanov, J.~G. K\"orner, {Quark model
  description of the tetraquark state $X(3872)$ in a relativistic constituent
  quark model with infrared confinement}, Phys. Rev. D 81 (2010) 114007.
\newblock \href {http://dx.doi.org/10.1103/PhysRevD.81.114007}
  {\path{doi:10.1103/PhysRevD.81.114007}}.

\bibitem{X3872asDD}
N.~A. Tornqvist, {Isospin breaking of the narrow charmonium state of Belle at
  3872-MeV as a deuson}, Phys. Lett. B590 (2004) 209--215.
\newblock \href {http://arxiv.org/abs/hep-ph/0402237}
  {\path{arXiv:hep-ph/0402237}}, \href
  {http://dx.doi.org/10.1016/j.physletb.2004.03.077}
  {\path{doi:10.1016/j.physletb.2004.03.077}}.

\bibitem{X3872asDD2}
E.~Braaten, M.~Kusunoki, {Exclusive production of the $X(3872)$ in $B$ meson
  decay}, Phys. Rev. D 71 (2005) 074005.
\newblock \href {http://dx.doi.org/10.1103/PhysRevD.71.074005}
  {\path{doi:10.1103/PhysRevD.71.074005}}.

\bibitem{X3872asMix}
R.~D. Matheus, F.~S. Navarra, M.~Nielsen, C.~M. Zanetti, {QCD sum rules for the
  $X(3872)$ as a mixed molecule-charmonium state}, Phys. Rev. D 80 (2009)
  056002.
\newblock \href {http://dx.doi.org/10.1103/PhysRevD.80.056002}
  {\path{doi:10.1103/PhysRevD.80.056002}}.

\bibitem{Ferreiro_comovers}
E.~Ferreiro, {Excited charmonium suppression in proton-nucleus collisions as a
  consequence of comovers}, Phys. Lett. B 749 (2015) 98.
\newblock \href
  {http://dx.doi.org/http://dx.doi.org/10.1016/j.physletb.2015.07.066}
  {\path{doi:http://dx.doi.org/10.1016/j.physletb.2015.07.066}}.

\bibitem{LHCb-DP-2008-001}
A.~A. Alves~Jr., et~al., {The LHCb detector at the LHC}, JINST
  3~(LHCb-DP-2008-001) (2008) S08005.
\newblock \href {http://dx.doi.org/10.1088/1748-0221/3/08/S08005}
  {\path{doi:10.1088/1748-0221/3/08/S08005}}.

\bibitem{LHCb-CONF-2019-005}
{LHCb collaboration}, {Multiplicity-dependent modification of $\chi_{c1}(3872)$
  and $\psi$(2S) production in $pp$ collisions at $\sqrt{s}= 8$
  TeV}~({LHCb-CONF-2019-005}).

\end{thebibliography}







\end{document}